\documentclass[useAMS,usenatbib,usegraphicx,twocolumn]{mn2e}

\title[Invariant manifolds and spiral arms in galaxies]
       {Invariant manifolds, phase correlations of chaotic orbits and
        the spiral structure of galaxies}

\author[N. Voglis, P. Tsoutsis, and C. Efthymiopoulos]
       {
       N. Voglis$^1$, P. Tsoutsis$^{1,2}$, and C. Efthymiopoulos$^1$\\
       $^1$Research Center for Astronomy and Applied Mathematics,
           Academy of Athens, Soranou Efessiou 4, GR-115 27 Athens, Greece\\
       $^2$Section of Astronomy, Astrophysics and Mechanics, Department of
           Physics, University of Athens,\\
           Panepistimiopolis, GR-157 84 Zografos, Athens, Greece\\
           e-mail: nvogl@academyofathens.gr, cefthim@academyofathens.gr,
           ptsoutsi@phys.uoa.gr
       }

\date{Released 2006 Xxxxx XX}
\pagerange{\pageref{firstpage}--\pageref{lastpage}} \pubyear{2006}

\def\LaTeX{L\kern-.36em\raise.3ex\hbox{a}\kern-.15em
    T\kern-.1667em\lower.7ex\hbox{E}\kern-.125emX}

\begin{document}
\label{firstpage}
\maketitle

\begin{abstract}
In the presence of a strong $m=2$ component in a rotating galaxy, the phase space
structure near corotation is shaped to a large extent by the {\it invariant
manifolds} of the short period family of unstable periodic orbits terminating
at L$_1$ or L$_2$. The main effect of
these manifolds is to create robust {\it phase correlations} among a number of
chaotic orbits large enough to support a {\it spiral} density wave outside
corotation. The phenomenon is described theoretically by soliton-like solutions
of a Sine-Gordon equation. Numerical examples are given in an N-Body simulation
of a barred spiral galaxy. In these examples, we demonstrate how the projection
of unstable manifolds in configuration space reproduces essentially the entire
observed bar-spiral pattern.
\end{abstract}


\section{Introduction}
Any gravitational theory of galactic spiral structure based on
orbits (see Contopoulos 2004, pp.497-503, for a review) must
provide a model of the way the {\it phases} of the orbits, i.e,
the angular positions of the apsides, become correlated, so as to
reproduce self-consistently the observed spiral pattern. Models
for normal spirals based on stable periodic orbits of the $x_1$
family, with apocenters aligned along the spiral, were proposed by
Contopoulos and Grosb{\o}l (1986, 1988). In such models, the
termination of the main spiral is placed near the 4/1 resonance
(Contopoulos 1985, Patsis et al. 1991, 1994, Patsis and Kaufmann
1999), while weak extensions may also be found beyond the 4/1
resonance, reaching the corotation region. On the other hand, such
models are probably not applicable at all if the spirals are
supported mostly by {\it chaotic} orbits, as, for example, in the
case when a bar is present (Kaufmann and Contopoulos 1996). The
dominance of chaotic orbits near and beyond corotation renders
these orbits `instrumental' (Kaufmann and Contopoulos 1996) in
successful self-consistent models of barred-spiral galaxies. In
particular, chaotic  orbits with Jacobi constants exceeding the
value at $L_{1,2}$ can circulate in and out of corotation, thus
supporting both the spiral and the bar (e.g. the `hot' population
of stars in the N-Body models of Sparke and Sellwood 1987).

A central question in the study of disk galaxies is whether the
bar drives spiral dynamics (e.g. Goldreich and Tremaine 1978,
Athanassoula 1980, Schwarz 1984, 1985), or the spiral structure is
a recurrent collective instability characterized by its own
independent dynamics (coupled perhaps to the bar dynamics, e.g.
Sellwood 2000 and references therein). In order to understand this
problem, a relevant key question is what {\it mechanisms} generate
the spiral pattern and whether such mechanisms work preferably
with regular or with chaotic orbits. In the present paper we
address the question of mechanisms generating phase correlations
of chaotic orbits able to support a spiral pattern.

The set of chaotic orbits near corotation fill stochastically a
connected chaotic domain of the phase space. The stochastic
character of the orbits could be naively perceived as opposed to
any persistent phase correlation among these orbits. However, it
is well known in the theory of dynamical systems that the
effective randomness exhibited by chaotic orbits is to a certain
extent only apparent. This is because of at least two reasons:

i) The loss of information along a chaotic orbit becomes important
only for times longer than the Lyapunov time of the orbit (the
inverse of the Lyapunov Characteristic Number).

ii) Even for times longer than the Lyapunov times the loss of
information in a chaotic domain, due to the exponential stretching
of a phase space volume element, which is maximum along one
particular local direction, is accompanied by a contraction of the
element across the same direction, i.e a gain of information
across this direction. Such local directions are determined by the
{\it invariant manifolds} of unstable periodic orbits in the
chaotic domain. The concept of invariant manifolds is a key
concept in understanding the chaotic dynamics in the corotation
region as will be analyzed in section 2 below.

In particular, we shall provide evidence that the unstable
invariant manifolds of the short-period unstable periodic orbits,
which form two families, one terminating at $L_1$ and the other at
$L_2$ (see section 2), provide a mechanism yielding phase
correlations of chaotic orbits that support, precisely, a spiral
density wave beyond the bar.We find that the intersections of
invariant manifolds with configuration space define patterns quite
similar to the spiral pattern of the observed density field. By
using an analysis parallel to the analysis given in Voglis (2003),
it is theoretically demonstrated (subsection 2.2) that the angular
positions of the apocenters of chaotic orbits with initial
conditions along an unstable manifold can be given, as a function
of time, by a soliton-type solution of a Sine-Gordon equation.
This solution is quantitatively accurate only for very small
amplitudes of the perturbation. However, even for large amplitudes
of the perturbation we find the same qualitative behavior, that is
the successive apocenters of chaotic orbits remain correlated for
quite long times, despite the positive Lyapunov exponents of these
orbits. This phase correlation causes the manifold to support the
spiral pattern. We furthermore provide a numerical calculation of
invariant manifolds in an N-Body simulation of a barred spiral
galaxy. In this calculation, we superpose the figures of the
manifolds on the locus of local maxima of the spiral density given
directly by the N-body particle distribution projected on the
plane of the disc. These two figures show a satisfactory
agreement.

A general property of invariant manifolds is that the unstable
manifold of one periodic orbit cannot intersect itself or the
unstable manifolds of other periodic orbits. When the phase space
is compact (i.e. no escapes are possible), the latter property
implies also that the manifolds of all the orbits fill densely the
phase space and that they are locally parallel to each other (the
same property holds also for stable manifolds). We find, with
numerical examples, that this geometric arrangement of the
manifolds is retained to a large extent even when the phase space
is not compact, i.e., some escapes are possible, provided that the
rate of escapes is slow. Given this property, we anticipate that
the effects shown below for a particular class of manifolds
corresponding to simple periodic orbits forming families
terminating at $L_1$ (or $L_2$) should in principle be present
also in the manifolds of many other families of unstable periodic
orbits. Finally, another important property of invariant manifolds
is their structural stability, i.e., the manifolds retain their
phase-space geometric structure under small perturbations of a
system.

The paper is organized as follows: In section 2 we define and
calculate the invariant manifolds of unstable periodic orbits in a
simple Hamiltonian model describing the dynamics in the corotation
region. In particular, we discuss the role played by invariant
manifolds in the dynamics of chaotic orbits. Section 3 presents
numerical results from an N-Body simulation of a barred spiral
galaxy. The manifolds are calculated first in a Poincar\'{e}
section and then their intersections with the 2D configuration
space of the disc are plotted. These plots are compared to the
spiral pattern of the system as derived by the particle
distribution on the disc. Section 4 summarizes the main
conclusions from the present study.

\section{Theory}

\subsection{The phase space structure near corotation and the form of
invariant manifolds}

The general form of the Hamiltonian of motion in the rotating frame of a disc
galaxy with a non-axisymmetric potential perturbation is:
\begin{equation}\label{hamgen}
H(r,\theta,p_r,p_{\theta})\equiv{1\over 2}\big(p_r^2+{p_\theta^2\over r^2}\big)
-\Omega_p p_\theta +V_0(r)+V_1(r,\theta)
\end{equation}
where $(r,\theta,p_r,p_\theta)$ are polar coordinates and their respective
conjugate canonical momenta, and $\Omega_p$ is the angular velocity of the
pattern. Assuming a bar-like perturbation $V_1$, a good first approximation to
the Hamiltonian (\ref{hamgen}) in the neighborhood of corotation is given by
(Contopoulos 1978):
\begin{equation}\label{hamcor}
H=\kappa_*I_1 + \alpha_*I_1^2 + 2b_*I_1I_2+c_*I_2^2 + A_*\cos
2\theta_2~~
\end{equation}
In this formula, $I_1$ is the action corresponding to epicyclic oscillations,
$\kappa_*$ being the respective epicyclic frequency. $I_2$ is an $O(A_*)$
correction to the quantity $p_\theta-J_*$, where $p_\theta$ is the angular
momentum of an orbit in the inertial frame and $J_*$ is the value of $p_\theta$
for a circular orbit at the corotation radius of the unperturbed $(V_1=0)$
potential. The angle $\theta_2$ is also an $O(A_*)$ correction to the azimuthal
angle $\theta$. For simplicity, in the analysis below we set
$I_2\approx p_\theta-J_*$ and $\theta_2\approx\theta$.

Since the epicyclic angle $\theta_1$, conjugate to $I_1$, is ignorable in the
Hamiltonian (\ref{hamcor}), this Hamiltonian model is integrable, $I_1$ being
a second integral independent of and in involution with $H$. However,  if any
other term $\exp(i(n\theta_1+m\theta_2))$ of the harmonic expansion of $V_1$
is added to the Hamiltonian (\ref{hamcor}), the Hamiltonian becomes, in general,
non-integrable and chaos is introduced to the model. As an example, we may
consider the Fourier modes $n=1,m=\pm 2$, which are important near Lindblad
resonances. A simple example of a Hamiltonian model including such terms is
\begin{eqnarray}\label{hamcper}
H=\kappa_*I_1 + \alpha_*I_1^2 + 2b_*I_1I_2+c_*I_2^2 + A_*\cos
2\theta_2~~~~~~~~~                             \nonumber\\
-\epsilon\big({2(I_1+I_{10})\over\kappa_*}\big)^{1/2}I_2
[\cos(\theta_1+2\theta_2) -\cos(\theta_1-2\theta_2)]
\end{eqnarray}
The dependence of the Fourier terms $\cos(\theta_1\pm 2\theta_2)$
on the actions $I_1,I_2$ introduced in Eq.(\ref{hamcper}) is a
simplified model used to demonstrate a number of phenomena
relevant to the discussion below, while, in reality, the
dependence of any Fourier term of $V_1$ on the actions would be
determined by the particular functional form of $V_1$. $I_{10}$ is
just a smoothing constant to avoid the singularity of the
derivative with respect to $I_1$.  At any rate, the crucial
parameter in all these phenomena is the overall size of the
perturbation terms $\cos(\theta_1\pm 2\theta_2)$, determined in
Eq.(\ref{hamcper}) by $\epsilon$.

For Jacobi constant values larger than the value at the unstable
Lagrangian equilibrium points $L_{1,2}$, there is a family of
unstable short-period orbits around $L_1$, or $L_2$ (Fig. 1,
schematic). These orbits form loops of size that increases as the
value of the Jacobi constant increases. We denote with $P_{L1,2}$
the point where a short-period orbit passes from its apocenter. We
shall also refer to the orbits themselves as the short period
orbits $P_{L1}$ (or $P_{L2}$). The curves with arrows correspond
to the equipotential passing from $L_{1,2}$ and the arrows
indicate the sense of flow of orbits inside corotation (prograde,
counterclockwise) or outside corotation (retrograde, clockwise). A
similar family of stable short-period orbits with apocenters
$P_{L4,5}$ exists also for the stable Lagrangian points $L_{4,5}$
for values of the Jacobi constant exceeding the value at
$L_{4,5}$.

Now, a visualization of the phase space of the system
(\ref{hamcper}) can be obtained by means of a suitable
Poincar\'{e} surface of section. Figure 2a shows the surface of
section $(I_2,\theta_2)$ corresponding to the successive crossings
of the angle $\theta_1$ along an orbit with the values $2n\pi$,
$n=1,2,\ldots$, i.e., at the apocenters of orbits, when $\epsilon$
is very small $(\epsilon=10^{-3})$. The basic phase portrait (Fig.
2a) resembles a pendulum phase portrait, with a separatrix
dividing the phase space into libration and rotation regions. The
fixed points $P_{L1,2}$ and $P_{L4,5}$ correspond to the same
points as in Fig. 1. The points $P_{L1,2}$ exist for a range of
values of the Jacobi constant $E_J>E_{J,L1,2}$, while the points
$P_{L4,5}$ exist for a range of values of $E_J$, such that
$E_J>E_{J,L4,5}>E_{J,L1,2}$. In the latter case, the phase
portrait is divided in two libration domains, around the stable
points $P_{L4,5}$, and two rotation domains, above and below the
unstable fixed points $P_{L1,2}$, in which the orbits make
rotations around the galaxy in either the retrograde sense (above
$P_{L1}$, outside corotation), or prograde sense (below $P_{L1}$,
inside corotation). On the other hand, if
$E_{J,L4,5}>E_J>E_{J,L1,2}$, there is an energetically forbidden
domain that replaces some of the KAM curves around $P_{L4}$ and
$P_{L5}$ of Fig. 2a.

Figure 2a gives initially the impression of being composed only by
invariant curves, i.e., by regular orbits. However, a closer look
reveals that there is a small degree of chaos in the neighborhood
of the unstable points $P_{L1,2}$. This chaos is clearly
manifested by computing the form of the {\it invariant manifolds}
emanating from $P_{L1}$ or $P_{L2}$ (Fig. 2b).

By definition, we call unstable manifold ${\cal W}^U$ of the short
period unstable periodic orbit $P_{L1}$ the set of points in
phase-space with pre-images approaching asymptotically the orbit
$P_{L1}$ as $t\rightarrow -\infty$. Similarly, the stable manifold
${\cal W}^S$ of $P_{L1}$ is the set of points in phase-space with
images approaching asymptotically the orbit $P_{L1}$ as
$t\rightarrow \infty$. Both sets, ${\cal W}^U$ and ${\cal W}^S$
are two-dimensional manifolds embedded in the three-dimensional
Jacobi constant hypersurface of the four-dimensional phase-space.

If, now, we work on a particular Poincar\'{e} surface of section,
such as, for example, the section of the apocenters as in fig.2,
the unstable periodic orbit $P_{L1}$ appears as an unstable fixed
point $P_{L1}$. Furthermore, the two-dimensional manifolds ${\cal
W}^U$ or ${\cal W}^S$ intersect this section along {\it
one}-dimensional submanifolds on the two-dimensional surface of
section. These submanifolds are the set of points approaching
asymptotically the fixed point $P_{L1}$ as $t\rightarrow -\infty$
or $t\rightarrow  \infty$, respectively. In order to distinguish
between two-dimensional manifolds in phase space and
one-dimensional manifolds in the surface of section, we call the
latter {\it apocentric} or {\it pericentric} manifolds, depending
on whether we take the surface of section at the apocenters or
pericenters of the orbits respectively. In fact, one can define a
surface of section in many different ways, but the definitions
adopted here, in terms of the apocenters or the pericenters, are
more convenient for the present study.

A well known theorem of dynamics (Grobmann 1959, Hartman 1960)
proves that:

a) in autonomous Hamiltonian systems the manifolds on the
Poincar\'{e} surface of section are invariant, i.e., coincide with
their images under the Poincar\'{e} map, and

b) they approach the fixed point $P_{L1}$ in the directions
tangent to the eigenvectors of the linearized map around $P_{L1}$.
In particular, the unstable manifold is tangent to the
eigenvectors associated with the absolutely larger real eigenvalue
(say $\lambda_1$) of the monodromy matrix at $P_{L1}$, while the
stable manifold is tangent to the eigenvectors associated with the
absolutely smaller eigenvalue, $\lambda_2$, of the same matrix. By
the symplecticity property of Poincar\'{e} maps we furthermore
have $\lambda_1\lambda_2=1$, thus $|\lambda_1|>1$ and
$|\lambda_2|<1$.

Property (b) can be exploited to calculate numerically the invariant manifolds
emanating from a particular unstable periodic orbit. Namely, let the functions
\begin{equation}\label{poinc}
\theta_2'=F(\theta_2,I_2),~~~ I_2'=G(\theta_2,I_2)
\end{equation}
represent the Poincar\'{e} map of the system under study (the functions $F$, $G$,
are determined numerically). We then calculate, by a Newton-Raphson scheme,
the position of one unstable periodic orbit, say $P_{L1}$, which corresponds to
one root $(\theta_{2,0},I_{2,0})$ of the set of equations $\theta_{2,0}=F(\theta_{2,0},
I_{2,0})$, $I_{2,0}=G(\theta_{2,0},I_{2,0})$. The monodromy matrix at the fixed point
$P_{L1}$ is defined as
\begin{equation}\label{monod}
A=\left(\begin{array}{cc}
{\partial F\over\partial\theta_2} &{\partial F\over\partial I_2} \\
{\partial G\over\partial\theta_2} &{\partial G\over\partial I_2} \\
\end{array}\right)_{(\theta_{2,0},I_{2,0})}
\end{equation}
where partial derivatives can also be determined numerically. The
eigenvalues $\lambda_1$ and $\lambda_2$,  of $A$, are real and
reciprocal. We then take a small initial segment $dS$ with one of
its endpoints coinciding with the fixed point $P_{L1}$ and the
other endpoint oriented on the direction of the eigenvector
associated with one of the eigenvalues $\lambda_1$, or
$\lambda_2$. Finally, we consider a large number of initial
conditions of orbits along the segment $dS$. The successive
Poincar\'{e} map images of the points along the segment associated
with $\lambda_1$ numerically determine the unstable
one-dimensional invariant manifold of $P_{L1}$. Similarly,
pre-images of points along the segment associated with $\lambda_2$
determine the stable one-dimensional invariant manifold of
$P_{L1}$.

Figure 2b shows a focus near $P_{L1,2}$ of the apocentric
invariant manifolds emanating from $P_{L1,2}$ for the same model
and parameter values as in Fig. 2a. The form of the manifolds of
$P_{L2}$, when they approach the neighborhood of $P_{L1}$, is
characteristic of what is known as `homoclinic chaos' in the
theory of Hamiltonian dynamical systems. Namely, the stable and
unstable manifolds have transverse intersections, called
homoclinic points. The part of a manifold between any two
successive homoclinic points determines a lobe. The successive
Poincar\'{e} images of the lobes are more and more elongated along
the two eigendirections, creating a very complicated homoclinic
tangle. That the complexity of the dynamics induced by the
homoclinic tangle is the source of chaos in Hamiltonian systems
was already pointed out by Poincar\'{e} (1892). However, a
non-schematic numerical calculation of invariant manifolds in a
simple Hamiltonian model was given relatively recently
(Contopoulos and Polymilis 1993). It has been established
theoretically that this dynamics is isomorphic to the dynamics of
the so-called $\Lambda-$~set (e.g. Contopoulos 2004, p. 152),
which obeys the Bernoulli shift map. The latter is considered as a
paradigm of chaotic system.

The geometric structure of invariant manifolds in phase space is
preserved even for large values of the nonlinearity parameter.
This effect is shown in Figs. 2c,d. The parameter $\epsilon$ has
now a value $10^{-2}$, i.e., ten times larger than in Figs. 2a,b.
As a result, the phase space near $P_{L1}$ is now largely chaotic,
and most rotational invariant curves are destroyed (Fig. 2c).
However, the apocentric invariant manifolds emanating from
$P_{L1,2}$ (Fig. 2d) preserve a geometric structure similar to
that of Fig. 2b, except for the fact that the overall vertical
scale of Fig. 2d is larger by a factor five from that of Fig. 2b.
This implies that, as $\epsilon$ increases, the size of the lobes
of the invariant manifolds increases. The lobes extend to bigger
distances away from $P_{L1,2}$.

Now, the {\it unstable} invariant manifolds of an unstable
periodic orbit, starting at the directions (U,UU), establish a
preferential direction of chaotic stretching in phase space in the
forward sense of time. In particular, all the orbits with initial
conditions inside the area defined by the lobes of one unstable
manifold are forced to follow the same preferential direction as
that of the manifold. Furthermore, the unstable manifolds of other
unstable periodic orbits in the same chaotic domain cannot
intersect the unstable manifold of the first orbit, thus they also
follow the same directions. Such directions are defined in the
neighborhood of every point in the chaotic domain and they serve
as the basis of construction of numerical indicators
distinguishing between ordered and chaotic orbits (Voglis et al.
1998, 1999, 2002).

Returning to the discussion of manifolds emanating from $P_{L1}$,
the relevant remark regarding the chaotic behavior of orbits near
corotation is that, as the lobes of the unstable manifold U, or
UU, approach close to $P_{L2}$, they become more and more
elongated, so that they repeatedly cross the straight lines
corresponding to the directions S, or SS (Figs. 2b,d). At every
such crossing, a lobe goes from the chaotic domain above $P_{L2}$
(outside corotation), to the domain below $P_{L2}$ (inside
corotation), or vice versa. As a result, all chaotic orbits with
initial conditions on, or inside, a lobe repeatedly go inside and
outside of the corotation region.

The sequence of these events has a Bernoulli shift dynamics, i.e., it is a very
chaotic phenomenon. This fact notwithstanding, we show in the next subsection that
the same dynamics imposed by unstable invariant manifolds implies that the {\it
phases} of chaotic orbits near an unstable manifold are organized along the
dominant directions of stretching, due to the area preservation, and they become
well correlated. This effect, precisely, serves as the basis of support of a
spiral density wave composed by chaotic orbits.

\subsection{Soliton-like motions and phase correlations of chaotic orbits}

Voglis (2003) has shown that in the neighborhood of {\it any} resonance in a disc
galaxy, the phases of stars with initial conditions close to the separatrix associated
with the resonance follow a {\it soliton}-type motion. A particular numerical
application was given in the case of the Inner Lindblad Resonance, but exactly the
same formalism can be applied to other resonances. In this section we shall
assume the Hamiltonian to be of the form (\ref{hamcper}) and describe
soliton-type correlations applicable to the orbits near $P_{L1,2}$.

To this end, we introduce the canonical transformation $(\theta_1,\theta_2,I_1,I_2)
\rightarrow (\theta_1',\theta_2',I_1',I_2')$ defined by
\begin{equation}\label{canon}
I_i={\partial F\over\partial\theta_i},~~~\theta_i'={\partial F\over\partial I_i'},~~~i=1,2
\end{equation}
where the generating function $F(\theta_i,I_i')$ is chosen in such a way as to
eliminate the dependence of the Hamiltonian on the angle $\theta_1$ up to terms of
order $O(\epsilon)$. Precisely, we have
\begin{eqnarray}\label{fchi}
F&=&\theta_1I_1'+\theta_2I_2'+\epsilon\big({2(I_1'+I_{10})\over\kappa_*}\big)
{I_2'\over\kappa_*}\times\\ \nonumber &
&[\sin(\theta_1+2\theta_2)-\sin(\theta_1-2\theta_2)]
\end{eqnarray}
Substituting the transformation (\ref{canon}) into the Hamiltonian
(\ref{hamcper}), the Hamiltonian takes the form:
\begin{equation}\label{hamcpern}
H=\kappa_*I_1' + \alpha_*I_1'^2 + 2b_*I_1'I_2'+c_*I_2'^2 +
A_*\cos 2\theta_2'+O(A\epsilon)
\end{equation}
where primed variables differ by $O(\epsilon)$ terms from their respective
non-primed variables, and both quantities $A$ and $\epsilon$ are assumed small.
Ignoring terms of order $O(A\epsilon)$, Hamilton's equations of motion in the
Hamiltonian (\ref{hamcpern}) can be cast in the form:
\begin{equation}\label{psieq}
\ddot{\psi}-\omega_0^2\sin\psi=0
\end{equation}
where $\psi=2(\theta_2'-\pi/2)$ and $\omega_0^2=-8c_*A_*$ (the constant $c_*$ is
negative, while $A_*$ is positive). The first integral of (\ref{psieq}) reads
\begin{equation}\label{psiint}
{\dot{\psi}^2\over 2} + \omega_0^2\cos\psi=C
\end{equation}
The quantity (\ref{psiint}) represents an approximate integral of
motion which, in a strict sense, is valid only for orbits lying on
invariant curves such as those of Fig. 2a. In particular, the
value of the constant $C=\omega_0^2$ corresponds to a theoretical
separatrix curve passing from $P_{L1,2}$. Along this curve, the
solution of (\ref{psieq}) reads
\begin{equation}\label{psitan}
\tan\big({\psi\over 4})=\exp[\pm(\omega_0t-k_0\xi)]
\end{equation}
where the parameter $\xi$ is related to the initial phase $\psi_0$ of an orbit
on the separatrix via $\xi=-{1\over k_0}\ln\tan(\psi_0/4)$, and
$k_0=\omega_0/\omega_1$,
with $\omega_1=\kappa_*+O(A_*)$. This choice of $\omega_1$ allows us to
follow the evolution of the phases $\psi$ for the successive Poincar\'{e}
consequents of the orbits when $\theta_1=\omega_1t=2\pi n$, with $n=1,2,\ldots$,
i.e., at the apocenters of the orbits.

Now, the solution (\ref{psitan}) can be viewed as a kink, or anti-kink, soliton-type
solution of the Sine-Gordon equation:
\begin{equation}\label{singor}
{\partial^2\psi\over\partial t^2}
+\omega_1^2{\partial^2\psi\over\partial\xi^2} - 2\omega_0^2\sin\psi
= 0
\end{equation}
In Eq.(\ref{singor}), $\xi$ was attributed the status of an independent variable of
the problem expressing the initial value of $\theta_2$ on the Poincar\'{e} surface
of section. The reason for this is that the number of stars in a galaxy is large
enough so as to expect that the separatrix defined by the apocenters of stars is
populated by stars all along its length. We may thus view Eq.(\ref{singor}) as
describing the evolution of the phases $\psi$ of a chain of stars populating the
separatrix. This is given as a function
of the time, $t$, but also of the initial phase of each star, given by $\xi$.

The above analysis was based on a canonical transformation of first order in the
perturation $\epsilon$, which also approximates parts of the manifolds as separatrices.
The question then arises whether the soliton-type flow of phases induced by
Eq. (\ref{singor}) is a good approximation of the true flow of phases along the
invariant manifolds as calculated numerically, i.e., without any approximation.
In order to explore this question, we first avoid the multiplicity of roots of
Eq.(\ref{psitan}) with respect to $\psi$ by differentiating this equation with
respect to time, that is
$\dot{\psi}=\pm 2\omega_0 sech(\omega_0t-k_0\xi)$, or
\begin{equation}\label{thdot}
\dot{\theta_2'}= \pm \omega_0 sech(\omega_0 t  -k_0\xi )
\end{equation}
In order to compare numerical results with analytical predictions, we then plot the
 theoretical relation (\ref{thdot}) against the numerical evaluation of $\dot{\theta}_2$
(by Hamilton's equations from the Hamiltonian (\ref{hamcper})) for
an ensemble of points being initially in a small segment along the
unstable eigenvector pointing in the UU direction, from $P_{L1}$,
and the U direction, from $P_{L2}$. As shown in Fig. 3, for a
relatively large value of the perturbation parameter
$\epsilon=10^{-2}$, and at different time snapshots, the numerical
solution $\dot{\theta}$ (displayed by dots in Fig. 3), as function
of $(\xi,t)$ follows approximately the analytical solution
(Eq.(\ref{thdot}), displayed by solid line in Fig. 3). In fact,
the theoretical solution for $\theta_2$ (Eq.\ref{psitan}) with
$\psi=2\theta_2'-\pi \simeq 2\theta_2-\pi$ has four
non-communicating branches which correspond to the
non-communicating parts of a theoretical separatrix that would
pass through $P_{L1,2}$(Fig. 2a) if chaos was absent, namely a)
upper and b) lower part for $0<\theta_2<\pi/2$, and
$3\pi/2<\theta_2<2\pi$ and c) upper and d) lower branch for
$\pi/2<\theta_2<3\pi/2$. These are merged to two branches for the
solution of $\dot{\theta}_2$ via Eq.(\ref{thdot}). On the
contrary, the invariant manifolds allow communication of the
branches via the expanding lobes as those shown in Fig. 2 b, d.
This means that when the orbits approach very close to $P_{L1,2}$
they are deflected to a different branch instead of approaching
asymptotically, as $t\rightarrow\infty$, $P_{L1}$ or $P_{L2}$.
This means that at, any time $t$, the motions of the manifold
points can be followed only within some limits $\xi_{min}\leq \xi
\leq \xi_{max}$ with $\xi_{max}-\xi_{min}$ large, but not infinite
as in the integrable case. Since particles jump to a different
branch of the equation (13), than the branch they were initially,
their phases $\theta_2$ are mixed in time. In order to circumvent
this problem, we redefine the value of $\xi$ for each orbit
whenever the orbit changes branch, according to Eq.(\ref{psitan}),
with $\psi=2\theta_2-\pi$ and $\theta_2$ equal to its value at the
particular moment when the particle changes branch.

The solitary motions shown in Fig. 3 correspond to a correlated
motion of the successive apocenters of chaotic orbits with initial
conditions along the manifold. This phenomenon, along with the
physical meaning of the changes of branches, will be discussed in
subsection 2.3 below. Furthermore, we find numerically that this
correlation persists to a large extent even for relatively large
values of the perturbation $\epsilon$ as a consequence of the
structural stability of the invariant manifolds. It is worth
examining the consequences of persisting phase correlations for
the dynamics induced by chaotic orbits in the corotation region.

\subsection{Mapping of the invariant manifolds in configuration
space and the form of spiral arms}

As described in subsection 2.1, the apocentric or pericentric manifolds
are one-dimensional objects embedded in a three dimensional space
of constant energy. This implies that they have one-dimensional
projections in any plane of this space. In the sequel, besides the
surface of section $(\theta_2,I_2)$, or $(\theta,p_\theta)$, we shall
consider the one-dimensional projections of manifolds in the physical,
or {\it configuration} space represented, e.g., by the canonical position
variables $(r,\theta)$, or $x=r\cos\theta$, $y=r\sin\theta$. This
allows to compare the topological characteristics of the manifolds
to morphological characteristics of the simulated galaxy.

In order to understand the geometric connection of the surface of
section to the configuration space, a key remark is that, in the
corotation region, the surface of section $(I_2,\theta_2)$ and the
configuration plane $(r,\theta)$, or $x=r\cos\theta$,
$y=r\sin\theta$, are nearly isomorphic. This isomorphy is shown
schematically in Fig. 4. The domains denoted A,B,C,D in Fig. 4a
are mapped to the domains A${^\prime}$,B${^\prime}$,C${^\prime}$,
and D${^\prime}$, respectively, in Fig. 4b. The mapping is derived
as follows: If the origin of epicyclic angles $\theta_1$ is
selected in such a way so that the Poincar\'{e} section points
$(I_2,\theta_2)$ for $\theta_1=2n\pi$, $n=1,2,\ldots$ correspond
to successive apocentric passages of an orbit, we have $\dot{r}=0$
at every section point. Replacing $p_\theta\simeq I_2+J_*$,
$\theta\simeq\theta_2$ in the general form of the Hamiltonian
(Eq.(\ref{hamgen})), the equation of the Jacobi constant
$H(r,\theta,p_r=0,p_\theta)=E_J$ may be solved for the variable
$r$, yielding $r$ at the apocenters as a function of
$\theta,p_\theta$ (or $I_2,\theta_2$) and of the constant $E_J$.
One readily finds that for potential functions corresponding to
the usual form of the rotation curve expected in a barred galaxy,
the apocentric radius $r$ is a monotonically increasing function
of $p_\theta$ both inside and outside corotation. In the inner
parts of the galaxy we expect that the velocity of circular orbits
raises as $v_c\propto r$ thus $r\propto p_\theta^{1/2}$. On the
other hand, beyond corotation the rotation curve is nearly flat up
to a distant radius, i.e., $v_c\sim const$, yielding $r\propto
p_\theta$. Finally, in the Keplerian limit $v_c\propto r^{-1/2}$,
thus $r\propto p_\theta^2$. These relations are not seriously
distorted by the inclusion of epicyclic oscillations. We shall see
in the next section that these monotonic relations are verified by
exact calculations in an N-Body system.

Because of the monotonic relation of $r$ and $p_\theta$, phase portraits
in the space $(p_\theta,\theta)$, or $(I_2,\theta_2)$, can be mapped
to nearly-isomorphic portraits in the space $(r,\theta)$, where $r$,
calculated as indicated above, always represents a local apocentric
distance of an orbit.

Figures 4a,b show the correspondence between the directions of the
stable and unstable manifolds emanating from $P_{L1,2}$ in the
section $(I_2,\theta_2)$ (Fig. 4a) and the same directions in
configuration space $(x,y)$ (Fig. 4b). In particular, the unstable
manifolds $U$, pointing upwards in Fig. 4a, are mapped to the
directions $U$ of Fig. 4b. In the separatrix limit, the unstable
manifold $U$ joins smoothly the stable manifold $S$, while the
unstable manifold $UU$ joins smoothly the stable manifold SS.
Furthermore, the branch US does not communicate with the branch
UUSS. We emphasize that such a theoretical separatrix can be
constructed for any value of the Jacobi constant at which the {\it
periodic orbit} $P_{L1}$ exists, i.e., the separatrix of Fig. 4b
should not be confused with an equipotential passing through the
{\it equilibrium point} $L_1$. The separatrix gives the slow drift
of apocenters along the branch US, or UUSS, that implies a slow
drift of the guiding centers of epicyclic motion for all orbits
with initial conditions on these two branches. These separatrices
do not coincide with the equipotential curves passing through
$L_1$, $L_2$ even in the limiting case when $P_{L1}$, $P_{L2}$
tend to $L_1$, $L_2$. In this respect, the manifolds shown in Fig.
4b are in a sense parallel to the `critical ergos curves' of
Lynden-Bell and Barot (2003), but at a higher value of the Jacobi
integral. By an elementary analysis of the linearized flow in the
neighborhood of the $P_{L1,2}$ orbits (named `Lyapunov orbits'),
Romero-Gomez et al. (2006) gave plots similar to Fig. 4b. In view
of the geometry of the separatrix, these authors identify the
manifolds as responsible for creating a {\it ring} structure in
barred galaxies like NGC1326. They also point out the robustness
of these objects, called by them `flux rings'.

In fact, as shown in the previous subsection, the robustness of
the phase flow along a manifold is a consequence of the
soliton-type motion of the phases of a chain of orbits populating
the manifold. This is shown in Figs. 4c,d, and 4e,f, which show
the invariant manifolds of Fig. 2d in phase space and in
configuration space after 4 iterations of an initial segment
$dS=10^{-4}$, (Figs. 4c,d) and after 12 iterations (Figs. 4e,f).
The correspondence in the form of manifold lobes in configuration
and phase space is clearly visible in these figures.

We emphasize that each point of figures 4d,f corresponds to a
position of apocenter of a chaotic orbit with initial conditions
along the manifolds. These positions change in time, for each
particular orbit, following the soliton-type flow described in
subsection 2.2. However, the pattern defined by the manifold
remains time-invariant. Furthermore, as the lobe oscillations
become of larger and larger amplitude, along the successive
iterations of the initial $dS$, the manifold moves far away from
the separatrix. Given that the sense of rotation of the galaxy is
counter-clockwise, we see that {\it the outward branches U of the
unstable manifolds emanating from $P_{L1,2}$ define trailing
spiral arms beyond $P_{L1,2}$}. This phenomenon will be studied in
detail in the next section. In the same time, the inward
directions UU support the bar, i.e., the same chaotic orbits with
initial conditions along the manifolds support both the spiral
pattern and the bar.

The opening of the manifold lobes is larger and faster when
$\epsilon$ increases. We shall see however, in the next section,
that the near-isomorphy between phase and configuration portraits
is preserved for quite large perturbations. In particular, in order
to study how far can extend the manifold spirals, we present a
numerical example referring to an N-Body simulation of a barred
spiral galaxy. It is then possible to show that {\it the invariant
manifolds U generate patterns that closely follow the form of the
N-Body spiral pattern of the system}.

\section{Numerical simulations: Comparison of invariant manifolds
and spiral arms}

Figure 5a shows one characteristic snapshot (for $t=47$ in units
of the half-mass crossing time) of an N-Body simulation of a
rotating barred - spiral galaxy. A series of such simulations with
$N=130000$, is presented in Voglis et al. (2006). These
experiments are produced with a smooth potential field code which
is an improved version of the code of Allen et al. (1990). The
initial conditions for a rotating system are produced from the
end-state of the so-called $Q-$model in the simulations of Voglis
et al. (2002). This is initially a triaxial configuration produced
by a collapse process with cosmological initial conditions
consistent with the standard $\Lambda-$CDM model. The initial
model has an almost zero spin parameter $\lambda$ (Peebles 1969).
In order to increase the total angular momentum of the model,
while respecting its energy and the scalar virial theorem, the
following velocity re-orientation process is implemented: At a
given snapshot of N-body evolution of the Q-model ($t_Q=100$), the
velocity component $v_{yz}$ of each particle, where the plane
$(y,z)$ is the plane of intermediate - longest axes, is
re-oriented so as to become perpendicular to the current
cylindrical radius $r_{yz}$ of the same particle. After the
re-orientation, the velocity components $v_{yz}$ of all particles
point to the direction of rotation clockwise with respect to the
$x-$~axis, which thus becomes the rotational axis of the system.
This is the way plots are presented in Voglis et al. 2006. In the
plots below, however, like Fig. 5a we have reversed the sense of
the x-axis so that the pattern in these figures is shown to rotate
{\it counterclockwise} (this is the most frequent choice in the
literature when discussing bar or spiral dynamics).

The so-produced system, called QR1, is let to evolve by the N-Body
code until a time $t_{QR1}=20$ (in units of the half-mass crossing
time), and then the same velocity re-orientation process is
implemented. The new system, QR2, is left to evolve again by the
N-Body code until $t_{QR2}=20$, when the velocity re-orientation
process is implemented once time. This process can yield different
rotating systems which have the same binding energy, and nearly
the same virial equilibrium, but they have consecutively higher
and higher amounts of angular momentum.

A detailed analysis of the regular and chaotic orbital content as well as of the
evolution of the above
systems is presented elsewhere (Voglis et al. 2006). We only mention here the results
relevant to the analysis below, which refer to the long-time evolution of the QR3
experiment, i.e., the experiment with initial conditions provided by three consecutive
applications of the velocity re-orientation algorithm:

a) The initially triaxial configuration has developed a thick disc
structure, with a thickness ratio of the order of 0.2. The
projection of the system on the disc plane (e.g. Fig. 5a) clearly
shows a bar as well as a grand design spiral pattern. In most
snapshots the calculated spiral pattern speed has nearly the same
value as the bar speed. The spiral amplitude, however, is
variable, ranging from a maximum 2 or 3 times smaller than the
amplitude of the bar to a minimum that is nevertheless always
above the threshold of statistical noise.

b) In some snapshots the spiral looks detached from the bar. Most snapshots,
however, manifest a continuity of the pattern between the bar and the spiral.

c) The orbits of all the particles are analyzed as regards their regular, or chaotic
character, by a combination of efficient numerical chaos indicators (Voglis et al.
2002). Let us note that the orbital analysis of a particular snapshot is possible
because the code provides an expansion of the potential in terms of a set of basis
functions, making thereby possible the explicit calculation of orbits, variational
equations etc. The main results are: i) The mass in chaotic motion is found to
be on the level of 60\% to 65\% of the total mass, and ii) when particles in
chaotic or ordered orbits are plotted separately, {\it the spiral arms are found
to be composed practically only by chaotic orbits}. On the contrary, the bar is
composed by both regular and chaotic orbits.

At the snapshot $t=47$, the bar-spiral structure is clearly seen
by simple visual inspection (Fig. 5a). All distances in this
figure were divided by the half mass radius of the system, which
is taken as the unit of length. The thick dots show the positions
of local maxima of the projected density on the $(y,z)$ plane.
This is done by splitting the distribution of points in concentric
rings $r,r+\Delta r$, where $\Delta r=0.2$, and calculating the
angles of local density maxima within each ring.

Figure 5b shows the distribution of the values of the Jacobi
constant for the set of particles which are located within small
projected area elements $r\Delta r\Delta\theta$, with
$\Delta\theta=2\pi/26$ around the values $(r,\theta)$
corresponding to the thick dots of Fig. 5a. The energy units are
derived by the numerical code length and time units, i.e., setting
the mass of each particle equal to $m_p=1$. In these units, the
half-mass crossing time is $t_{hmct}=1.1085\times 10^{-4}$, the
half-mass radius is $R_h=9.26 \times 10^{-2}$. The root-mean
square velocity is found to be $v_{rms}=732.6$. The pattern speed
(estimated numerically by the rate of angular displacement of the
density maxima of the pattern), is equal to $\Omega_p=2\pi/9.43$
in units of $1/t_{hmct}$, i.e., the period of pattern rotation is
about ten times larger than the half-mass crossing time (which
corresponds to a typical period of the particles' orbits). This
means that the rotating pattern can be safely characterized as an
evolving density wave.

Another approximation in the calculation of Fig. 5b is the fact
that the kinetic energy of a particle includes a $v_x^2/2$ term,
which corresponds to the kinetic energy of the velocity component
perpendicularly to the disc. This small term (a few percent of the
total kinetic energy) is substracted from the evaluation of the
Jacobi constant, since, for most particles, the motion
perpendicularly to the disc is effectively decoupled from the
motion on the disc plane. We thus approximate the Jacobi constant
for each particle by:
\begin{equation}\label{jacnb}
E_J={1\over 2}(v_y^2+v_z^2)-\Omega_p(zv_y-yv_z) + V(x,y,z)
\end{equation}
where the function $V(x,y,z)$ is calculated by the series of basis
functions of the N-Body code.

Figure 5b shows the distribution of Jacobi constants, with the
above conventions, for the particles located in the maxima shown
in Fig. 5a, where, in addition, we only count the particles in
rings beyond the value $r=1$, marking, approximately, the end of
the bar. Thus, the distribution of Fig. 5b represents essentially
the distribution of values of the Jacobi constant for particles
located along the spiral arms. The peak of this distribution is at
a value $E_J=-1.12\times 10^6$, which is slightly above the value,
$E_{J,L1}=-1.133\times 10^6$.

As described in subsection 2.1, the apocentric manifolds are
one-dimensional manifolds in the three-dimensional subspace ($r,
\theta, p_{\theta}$) of the entire four-dimensional phase space of
the system. Figure 6 shows a calculation of the unstable
apocentric manifolds emanating from the periodic orbit $P_{L2}$ in
the potential of the system of Fig. 5a (with initial conditions
$\theta=3.2247$, $p_\theta=1.194$). Three different projections of
these manifolds are shown, namely a) on the surface of section
$(\theta,p_\theta)$, $\dot{r}=0$, $\ddot{r}<0$ (Fig. 6a), b) on
the plane $(r,p_{\theta})$ (Fig. 6b) and c) on the configuration
space $(r, \theta)$  (Fig. 6c). The Jacobi constant has the value
$E_J=-1.125 \times 10^6$, close to the maximum of the distribution
of Fig. 5b. (We only examine theoretical orbits with $x=0$, i.e.,
lying on the disc plane). Fig. 6 shows the seventh Poincar\'{e}
map image of two initial segments of length $dS=10^{-4}$ on the
Poincar\'{e} section, tangent to the unstable directions U and UU
respectively, at the fixed point $P_{L2}$. As a property of the
manifold, the seventh image of the initial $dS$ contains also all
the previous images from the first to the seventh. This choice of
initial length was made by the criterion that the numerically
calculated direction of the segment after its first iteration
coincides with the eigendirection as calculated by the monodromy
matrix of the periodic orbit $P_{L2}$ up to 3 significant digits.
Furthermore, each segment is populated with 10000 equally spaced
points of initial conditions. Figure 6c shows the calculated
projections of the apocentric manifolds U and UU (2$\times$10000
points) as well as their symmetric manifolds with respect to the
center $y=z=0$, which are the manifolds emanating from $P_{L1}$.

Figure 6a is different from the simple picture of manifolds shown
in Figs. 2b,d in many aspects. A first difference is that the
$\theta$ value of $P_{L2}$ is close to the value $\theta=\pi$,
i.e., shifted by $\pi/2$ with respect to Figs. 2b,d. This is
because at the snapshot $t=47$, the bar is oriented almost
horizontally (Fig. 5a). But the most serious difference is that
the manifolds of Fig. 6a are broken into disjoint pieces. This
breaking of the image of the manifold on the surface of section is
an artifact caused by the particular choice of surface of section.
This can be exemplified with the help of Figs. 7a,b,c,d which
refer to the same manifold, but calculated for six iterations of
the initial $dS$, in order to obtain a clearer figure. Figures
7c,d show the time evolution of two nearby orbits one of which
gives the last point, P, of a disjoint piece of the manifold of
Fig. 7a, while the other orbit gives the first point $P'$ of the
next disjoint piece of the manifold in the same figure. We notice
that while the orbits are really nearby (Fig. 7d), the first orbit
has a local maximum of $r(t)$ at the point $P$, while the second
orbit has no local maximum of $r(t)$ at times near the time of the
$P-$maximum. In fact, the second orbit yields a local maximum much
later, at the point $P'$ (Fig. 7d), thus breaking the continuity
of Poincar\'{e} intersections in Fig. 7a between the points $P$
and $P'$. The same phenomenon is also shown in Fig. 7c, where we
clearly see that, while the first orbit on the surface of section
($\dot{r}=0$,$\ddot{r}<0$), represented by the line $\dot{r}=0$ in
Fig.7c, intersects the line $\dot{r}=0$ at P, the second orbit
passes close to the section line, but without intersecting this
line.

We emphasize that the above effect is an artifact of the
particular choice of surface of section, which is not a true
Poincar\'{e} section (Poincar\'{e} 1892). In fact, it is hardly
possible to find any surface of phase space in the above model
that can be {\it proven} to be a true Poincar\'{e} section. Even
so, the invariant manifold surfaces {\it are} continuous in phase
space, and it is only their intersections with the particular
surface of section which are discontinuous. Furthermore, the basic
topological property of invariant manifolds, namely the fact that
they cannot intersect themselves, is also preserved in the surface
of section images of the manifold shown in Figs. 6a or 7a. This
fact implies that, as the number of iterations increases, i.e.,
when the manifold is calculated for longer times, the manifolds
fill the connected chaotic domain embedding the fixed points
$P_{L1,2}$ in the surface of section. This tendency is clearly
observed by comparing Fig. 6a, which depicts the manifold after
only one more iteration than Fig.7a. In Fig.6a the manifold
crosses a larger area in the chaotic domain and it starts showing
the effect discussed in section 2, i.e., it defines {\it
preferential directions} in the surface of section, given by many
manifold segments forming narrow bundles in almost parallel
directions. Such bundles are clearly visible in both Figs. 6a and
7a.

Let us return to the three projections of the apocentric manifolds
on the three planes $(\theta,p_\theta)$, $(r,p_{\theta})$, $(r,
\theta)$ shown in Figs. 6 a, b, c, respectively. The projection of
the apocentric manifolds on the plane $(r,p_{\theta})$ shows an
impressively narrow correlation between $r$ and $p_{\theta}$. For
every branch (U, or UU) there is a monotonic (on average)
dependence of $r$ on $p_{\theta}$. This is a numerical
verification, in the case of the N-Body experiment, of the
monotonic relation between $r$ and $p_{\theta}$, that was
discussed in subsection 2.3.

A direct result of this monotonic (on average) dependence (of the
apocentric distances $r$ and the angular momentum $p_{\theta}$ of
the orbits on the invariant manifolds) is that causes a
near-isomorphy between the projection of apocentric manifolds in
phase space $(\theta,p_\theta)$ (shown in Fig. 6a) and in the
configuration space $(r, \theta)$ (shown in Fig. 6c).

A main effect caused by this near-isomorphy is that the bundles of
preferential directions of the apocentric manifolds that appear as
almost straight lines in Fig. 6a, appear as spiral lines in Figs.
6c. In other words {\it the loci of preferential directions
defined by the apocentric manifolds define a geometric pattern on
the configuration space that, except for some details, has the
general form of a grand design spiral.}

It should be stressed that all the points of the apocentric
manifolds in Figs. 6, 7a,b correspond to local apocentric passages
of the respective orbits. Thus, one single orbit does not describe
continually in time the manifolds shown in these figures. But if
we take a chain of stars that densely populate the manifolds, this
chain defines a flow along the manifold that keeps the apocenters
of chaotic orbits correlated for quite long times. This
correlation was modelled theoretically in subsection (2.2) as a
soliton-type flow of the successive positions of the apocenters of
chaotic orbits.

In Fig. 8 the positions of maxima of the particle distribution
along the N-body spiral arms (Fig. 8a, same as Fig. 5a) are
superposed on the spiral patterns generated by the preferential
directions of the unstable invariant manifolds of the $P_{L1,2}$
family (plotted for three different values of the Jacobi constant
in Figs. 8b, c, d, which are in a narrow band of values larger
than the Jacobi constant at the Lagrangian points $L_1$,$L_2$, and
close to the peak of the distribution of the energies of Fig. 5b).
It is immediately seen that the manifold spirals are very closely
aligned to the real spiral arms. The images of the apocentric
manifolds on the configuration space, for the three different
values of $E_J$, are similar to each other, implying that their
superposition enhances the overall spiral pattern induced by the
manifolds.

The fact that the projections of apocentric manifolds in the
configuration space are well-correlated with the maxima of the
spiral density is remarkable, because the form of the manifolds
reflects topological properties of the underlying chaotic phase
space, while the density maxima are related  to the particles'
(chaotic) motion in configuration space. We can make a number of
comments pointing to the theoretical understanding of this
phenomenon:

i) We have compared the distribution of all the particles in
chaotic orbits of the system along the Jacobi constant axis given
in Voglis et al. 2006 (Fig.15 therein) with the distribution of
the particles shown in Fig. 5b. The two distributions are quite
similar, and their maxima are projected at about the same value of
$E_J$, which is very close to the band of values of $E_J$ for
which the manifolds of Fig. 8 are calculated. This means that the
backbone of the spiral arms is related to chaotic orbits with
initial conditions close to the apocentric manifolds of figure 8.
Particles with values of the Jacobi integrals much different than
those near the maximum of the distribution of Fig. 5b partly
contribute to the density along the spiral arms. However, we have
checked that most of these particles contribute to the density of
the axisymmetric background.

ii) Near the bundles of preferential directions, the manifolds
act, across these directions, as {\it attractors} of the orbits,
since any element of an area-preserving phase flow has to contract
in the direction across the manifolds in order to compensate for
the strong stretching along the manifolds. In fact, the stretching
and twist properties of the manifolds are far from being uniform
along their length. The attraction across the manifold, as well as
the organization of phases, is stronger where the stretching along
the manifold is also stronger. This mechanism is more effective
along bundles of preferential directions, while it is less
effective near the turning points of the manifold. These phenomena
are similar to the `stickiness' phenomena caused by invariant
manifolds in model Hamiltonian dynamical systems (e.g.
Efthymiopoulos et al. 1997, Contopoulos et al. 1999).

iii) Finally, another property of the manifolds is that they are
{\it recurrent} (Contopoulos and Polymilis 1993), i.e., the
manifold lobes return many times near the points $P_{L_{1,2}}$
even if they temporarily go to large excursions away from
$P_{L_{1,2}}$. This mechanism enhances further the density along
the segments of strong stretching, and also agrees with the maxima
of the density appearing preferably along bundles of the manifold.
The recurrence of the manifolds is particularly important in the
system under study, because the phase-space is non-compact and the
recurrence of chaotic orbits cannot be established by the
Poincar\'{e} recurrence theorem which is applicable only to
compact systems. It is remarkable that the recurrence of the
manifolds can create new stickiness effects (recurrent
stickiness), leading to further accumulation of points in the
sticky region (Contopoulos and Harsoula 2006).

Figure 9 shows a comparison of invariant apocentric manifolds with
real spirals at three different time snapshots of the evolution of
the N-Body experiment, namely $t=18$ (Figs. 9a,b), $t=47$ (Figs.
9c,d) and $t=74$ (Figs. 9e,f). (The manifolds are plotted after
eight iterations, when the spread of numerical points along the
manifold is still small enough to allow for a coherent
visualization of the manifold).

The main difference in the system at these three snapshots is
that, as $t$ increases, the pattern speed $\Omega_p$ of the bar
has a tendency to decrease as a result of torques exerted to the
bar by the rest of the system and the angular momentum
transference outwards. The decrease of $\Omega_p$ is significant,
of order 20\%, between $t=18$ ($\Omega_p=0.8037$) and $t=47$
($\Omega_p=0.6662$), while it is less important (about 6\%)
between $t=47$ and $t=74$ ($\Omega_p=0.6282$).

As the system slows down, the position of $L_{1,2}$ moves outwards
and the Jacobi constant $E_{J,L_{1,2}}$ increases. The unstable
manifolds in Figs. 9b,d,f are calculated for the periodic orbit of
the corresponding $P_{L1,2}$ family at the Jacobi constant values
$E_J=-1.25\times 10^6$, $E_J=-1.131\times 10^6$ and
$E_J=-1.085\times 10^6$, respectively. These values are slightly
larger than the Jacobi constant values at the respective
Lagrangian points $L_{1,2}$, namely $E_J=-1.26\times 10^6$,
$E_J=-1.133\times 10^6$ and $E_J=-1.088\times 10^6$.

Clearly, as the pattern speed of the bar decreases, the manifolds
become more and more open, that is they extend to larger distances
beyond corotation under the same number of iterations.
Furthermore, Figs. 9b,d,f show that the agreement of the
apocentric manifold with the spiral pattern is good precisely in
the regions where the manifold segments accumulate in bundles of
preferential directions. The overall connection of the manifold's
spiral geometries is generally visible in all three figures,
except for a weak spiral extension in Fig. 9a corresponding to the
outermost points of the spiral maxima of Fig. 9b (this is probably
a transient effect).

We conclude that as the manifolds become more and more open, the
bundles of preferential directions become broader, but also the real
spiral becomes weaker and less clearly defined. A detailed explanation
of these phenomena is necessary, since they all point to the direction
of a clear connection between the dynamics induced by invariant manifolds
and the dynamics of real spiral arms. In particular, given that the locus
of apocenters defined by the manifolds is the place where stars spend a
large part of their radial period, we conclude that the projections of
the apocentric manifolds reveal a narrow dynamical connection between
manifolds and spiral arms.

Finally, Fig. 10 shows a calculation of a number 300 of individual
stellar orbits (plus another 300 symmetric orbits) with initial
conditions along the unstable direction of the manifolds of
Fig.8b. The twelve different panels correspond to twelve
successive positions of the pericenters (white thick dots) and
apocenters (black thick dots) of these orbits. The first panel
(left-top) is when the apocenters start leaving the neighborhood
of $P_{L1,L2}$. This figure summarizes the previous conclusions
from the point of view of the orbital flow. Namely, we observe
that the phases of these orbits remain very well correlated for
nine radial periods, and to some extent for another three periods.
Furthermore, apocenters are preferably aligned along the spiral
pattern, while pericenters are located mostly in a double
ring-like locus surrounding the bar and the stable Lagrangian
points $L_4$, $L_5$. It is remarkable that inside the corotation
radius pericenters and apocenters cooperate to support a shell of
chaotic orbits surrounding the bar. Outside the corotation radius,
for some azimuthal extent from the Lagrangian points $L_1$, $L_2$,
pericenters and apocenters also cooperate to form an enhanced
initial arc of the spiral arm. For larger azimuthal angles,
however, the pericenters form an arc surrounding the stable
Lagrangian points $L_4$, $L_5$, while the apocenters form a more
open arc following the spiral arms.

Thus, we conclude that the apocentric manifolds are more important
than pericentric manifolds, for the formation of spiral arms, not
only because of the longer time intervals spent by the orbits near
the apocenters than near the pericenters, but also because only
the apocentric manifold is responsible for the extension of the
spiral arms to large distances from the corotation radius.

\section{Conclusions}
We have explored the role of the invariant manifolds emanating
from simple unstable periodic orbits (forming families terminating
at $L_1$ ,or $L_2$) in producing a mechanism of phase correlations
between chaotic orbits beyond corotation in a barred spiral
galaxy. In particular, we define the apocentric (or pericentric)
one-dimensional manifolds as the intersections of the phase space
manifolds by the surface of section of the apocenters (or the
pericenters). A theoretical analysis is made, based on a theory of
soliton-type motions of phases (Voglis 2003) of orbits with
initial conditions along the apocentric manifolds. A numerical
analysis is also carried out, referring to the calculation of
invariant apocentric manifolds in the potential of an N-Body
simulation of a barred spiral galaxy. In this case we explore the
connection between the manifolds and the overall dynamical
features of the N-Body system. Our main conclusions are as
follows:

1) While every orbit with initial conditions on the unstable
apocentric manifolds of the $P_{L1,2}$ unstable periodic orbits is
chaotic, the phases (the positions of the apocenters) for a chain
of particles with initial conditions populating the manifold
remain correlated for quite long times. This correlation can be
approximated by a soliton-type motion for the phases. This is true
not only for small amplitude of the perturbation, but also, to a
large extent, for large amplitude of the perturbation.

2) The portraits of phase space, as described by the apocentric
manifolds, have nearly-isomorphic portraits into configuration
space. These portraits are described in subsection 2.3 as well as
in Fig.6. In particular, for small values of the perturbation, the
apocentric manifolds form a double-ring structure surrounding the
bar and the stable Lagrangian points $L_4$,$L_5$. But when the
perturbation is strong enough, the images of the manifolds in
configuration space extend to large distances beyond corotation,
and they take the form of trailing spiral arms.

3) A comparison of unstable invariant apocentric manifolds with
the density maxima of an N-Body experiment along the spiral shows
an overall connection of the manifold spirals with the N-Body
spirals. In particular, the two sets are closely aligned in all
regions where the manifold segments form narrow bundles along
preferential directions of strong stretching.

4) As long as the galaxy's pattern slows down, the manifolds
become gradually more open and the bundles of preferential
directions become broader, but still the connection between
manifold and spiral geometry is preserved. On the other hand, the
amplitude of the spiral becomes also weaker, and its shape less
clearly defined. In conclusion, we find a clear dynamical
connection between the loci of apocenters defined by the manifolds
of unstable periodic orbits and spiral arms. In this respect, the
apocentric manifold is narrowly associated with the positions of
maxima of the density along the spiral arms. Furthermore, this
connection is preserved for an important fraction of the simulated
galaxies' lifetimes.

As a final remark, we may revisit the question posed in the
introduction on whether the above mechanism can help clarify the
role of the bar in driving the dynamics of spiral arms. Our
findings suggest that, through the manifolds, the bar can indeed
initiate a spiral structure. Notice that the unstable manifolds
are asymmetric in configuration space with respect to the bar axis
(fig.4) and, as shown in section 2, they can initiate trailing
spirals, provided that the bar is strong and/or rotates fast
enough to create a large effective perturbation near corotation.
Furthermore, this mechanism exploits the relative flexibility of
chaotic orbits to exhibit large excursions in phase space while
remaining correlated as regards their phases. On the other hand, a
spiral pattern initiated by a bar through the above mechanism
amplifies the effective perturbation even more, a fact that acts
additively to the whole phenomenon, which is based on chaotic
orbits. These points, which may prove to be crucial for the
longevity of the induced spiral pattern, are a subject of current
research.

{\bf Acknowledgements:} This project was supported in part by the
Empeirikion Foundation. Stimulating discussions, as well as a very
careful reading of the manuscript by Prof. G. Contopoulos are
gratefully acknowledged.


\clearpage


\onecolumn
\begin{figure}
\centering
\includegraphics[width=12cm]{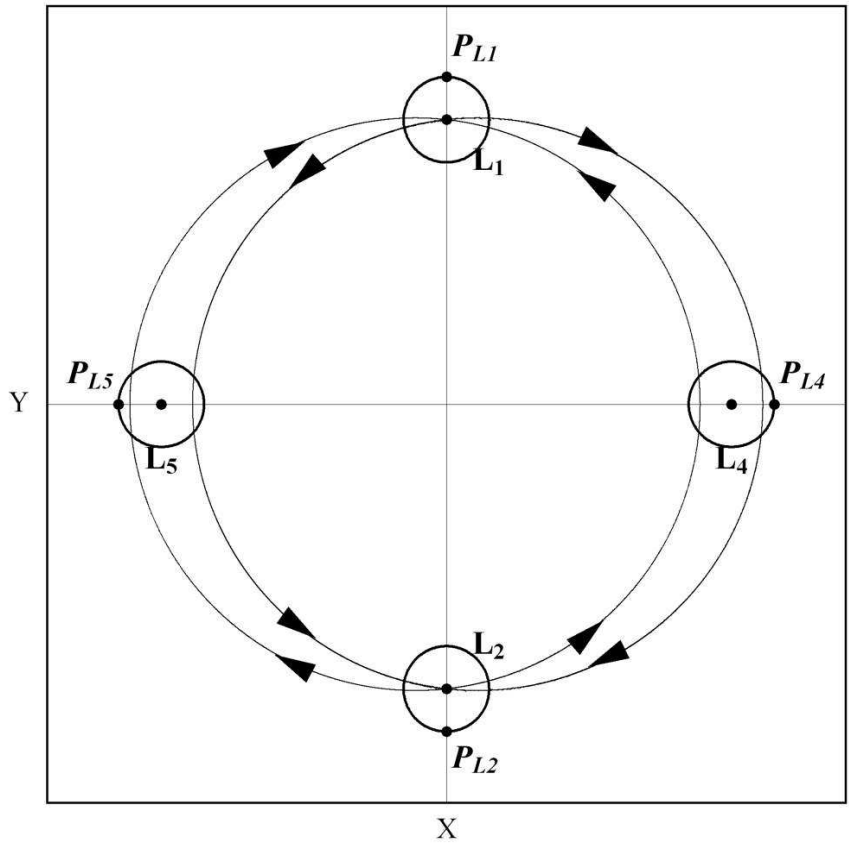}
\caption{Short period orbits around the Langrangian equilibria
$L_{1,2}$ (unstable) and $L_{4,5}$ (stable). The curves with arrow
correspond to the equipotential passing from $L_{1,2}$. The arrows
indicate the sense of rotation inside and outside corotation.}
\label{figure2}
\end{figure}
\clearpage

\begin{figure}
\centering
\includegraphics[width=\textwidth]{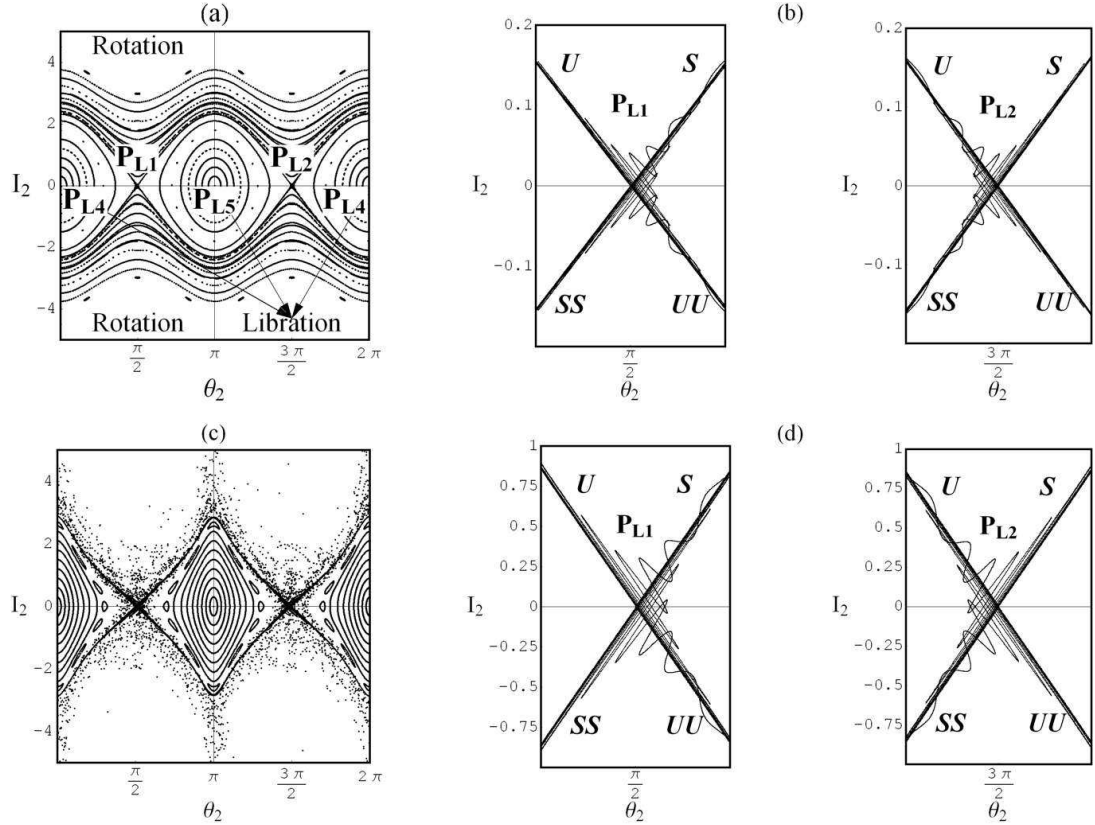}
\caption{The Poincar\'{e} surface of section $(I_2,\theta_2)$ for
$\theta_1=2k\pi, k=0,1,2,\ldots$ in the Hamiltonian (\ref{hamcper})
with $\kappa_*=3.23414\times 10^{-1}$, $\alpha_*=-2.90577\times
10^{-2}$, $b_*=0$, $c_*=-7.5\times 10^{-3}$, $A_*=1.897\times
10^{-2}$ and (a) $\epsilon=10^{-3}$, (c) $\epsilon=10^{-2}$. In both
cases the value of the Jacobi constant is set equal to
$E_J=-1.897\times 10^{-2}$. The stable and unstable manifolds of the
periodic orbits $P_{L1,2}$ (unstable fixed points in (a),(c)) are
shown in (b) and (d) respectively. } \label{figure2}
\end{figure}

\begin{figure}
\centering
\includegraphics[width=\textwidth]{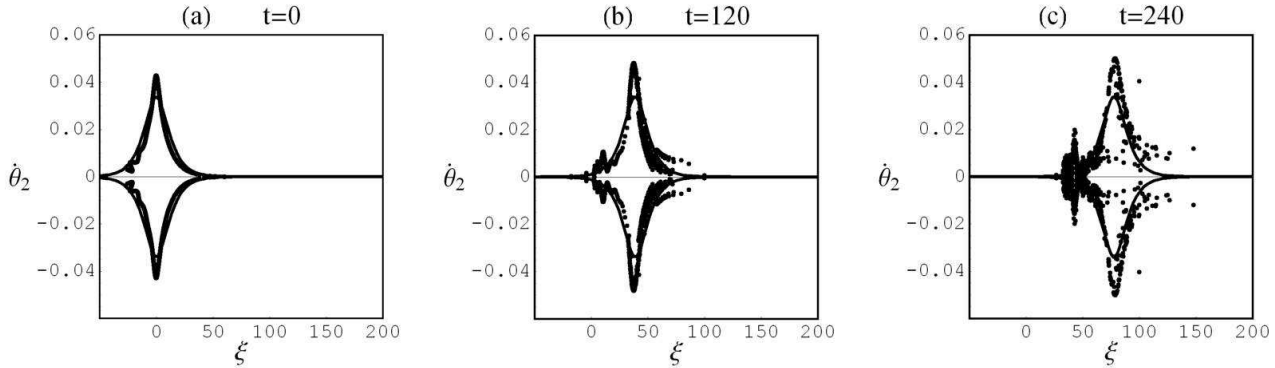}
\caption{Comparison of the theoretical soliton-type solution given
by Eq.(\ref{thdot}),(solid line) with the corresponding numerical
relation for 200 points along the unstable manifold U emanating
from $P_{L1}$ in the Hamiltonian (\ref{hamcper}) with parameters
as in Fig. 2d. The different panels correspond to different time
snapshots, namely a)$t=0$, b) $t=120$, c) $t=240$.}
\label{figure3}
\end{figure}

\begin{figure}
\centering
\includegraphics[width=14cm]{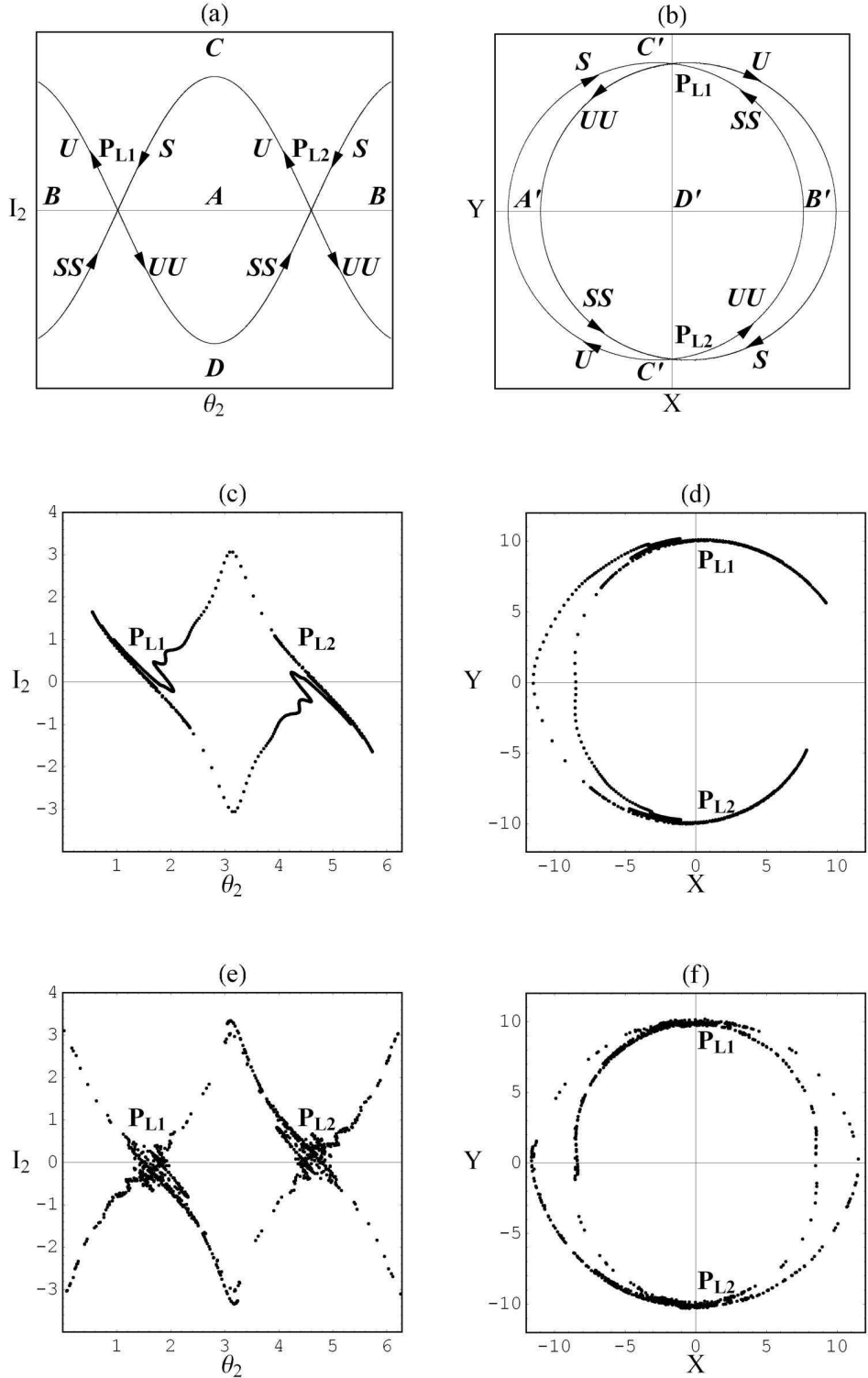}
\caption{Schematic representation of the isomorphy between a) the
phase portrait in the Poincar\'{e} section $(I_2,\theta_2)$, and
b) configuration space. The regions A,B,C,D in (a) correspond to
A$^\prime$, B$^\prime$, C$^\prime$, D$^\prime$ in (b). The arrows
indicate the directions of the stable and unstable manifolds
emanating from $P_{L1,2}$ in phase space and configuration space.
The manifolds emanating from $P_{L1,2}$ are shown in phase space
(c,e), and configuration space (d,f) for a short time (four
iterations of an initial segment $ds=10^{-4}$), (c,d) and longer
time (12 iterations), (e,f).} \label{figure4}
\end{figure}

\begin{figure}
\centering
\includegraphics[width=\textwidth]{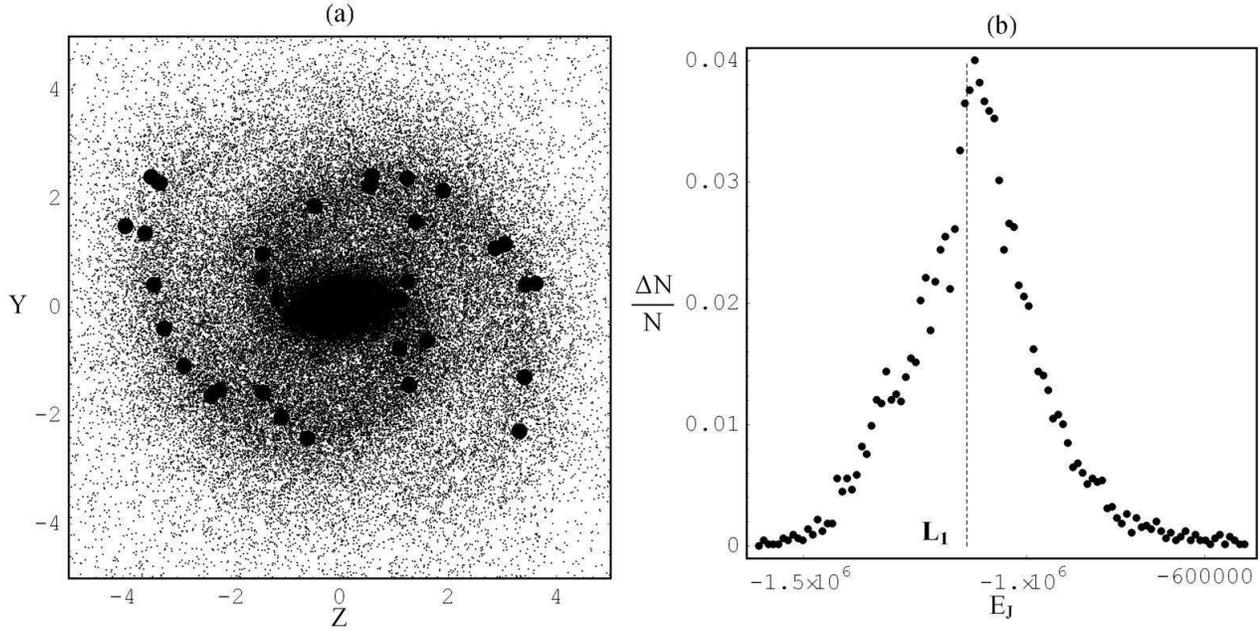}
\caption{ a) Projection of the particles of the N-Body simulation
described in section 3 on the $(y,z)$ plane, at the time snapshot
$t=47$. In this projection the galaxy rotates counterclockwise.
The thick dots are the maxima of the surface density calculated
with a division of the system in concentric rings (see text for
details). b) The distribution of the Jacobi constant values (in
units of the N-body code) for the particles located in small bins
$\Delta r\Delta\theta$ centered at the thick dots of (a). The
dashed vertical line indicates the Jacobi constant value for
$L_1$.} \label{figure5}
\end{figure}

\begin{figure}
\centering
\includegraphics[width=\textwidth]{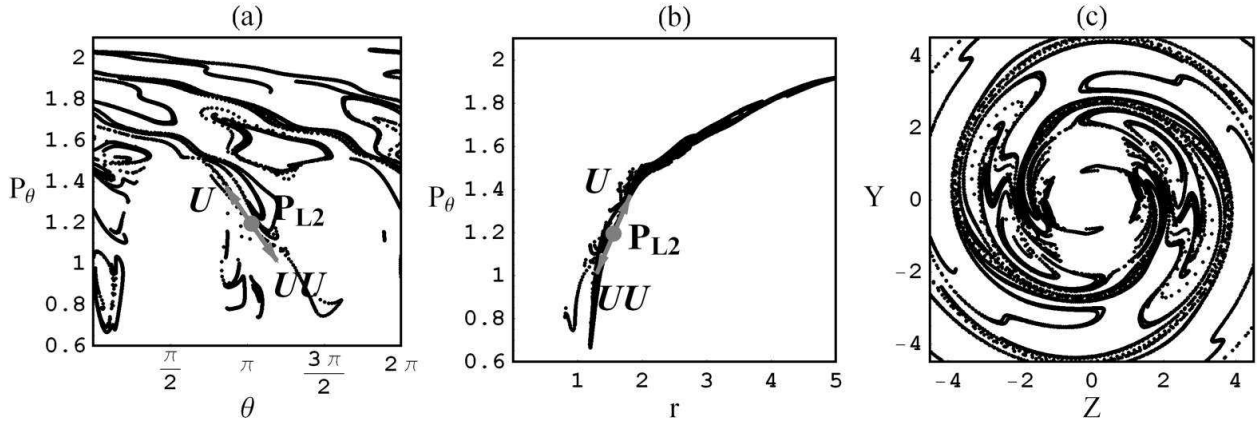}
\caption{Projections of the apocentric manifolds: (a) on the
surface of section plane $(\theta, p_{\theta})$, (b) on the plane
$(r, p_{\theta})$ and (c) on the configuration space $(r, \theta)$
or $(z,y)$. In (a) and (b) the unstable fixed point $P_{L2}$ and
the corresponding initial directions of the manifolds U and UU are
also shown. In (c) the projection of the symmetric manifolds,
emanating from the fixed point $P_{L1}$, is shown together with
the manifolds from $P_{L1}$. The monotonic (on average) dependence
of $r$ on $p_{\theta}$ shown in (b) produces a near-isomorphy
between (a) and (c).} \label{figure6}
\end{figure}


\begin{figure}
\centering
\includegraphics[width=\textwidth]{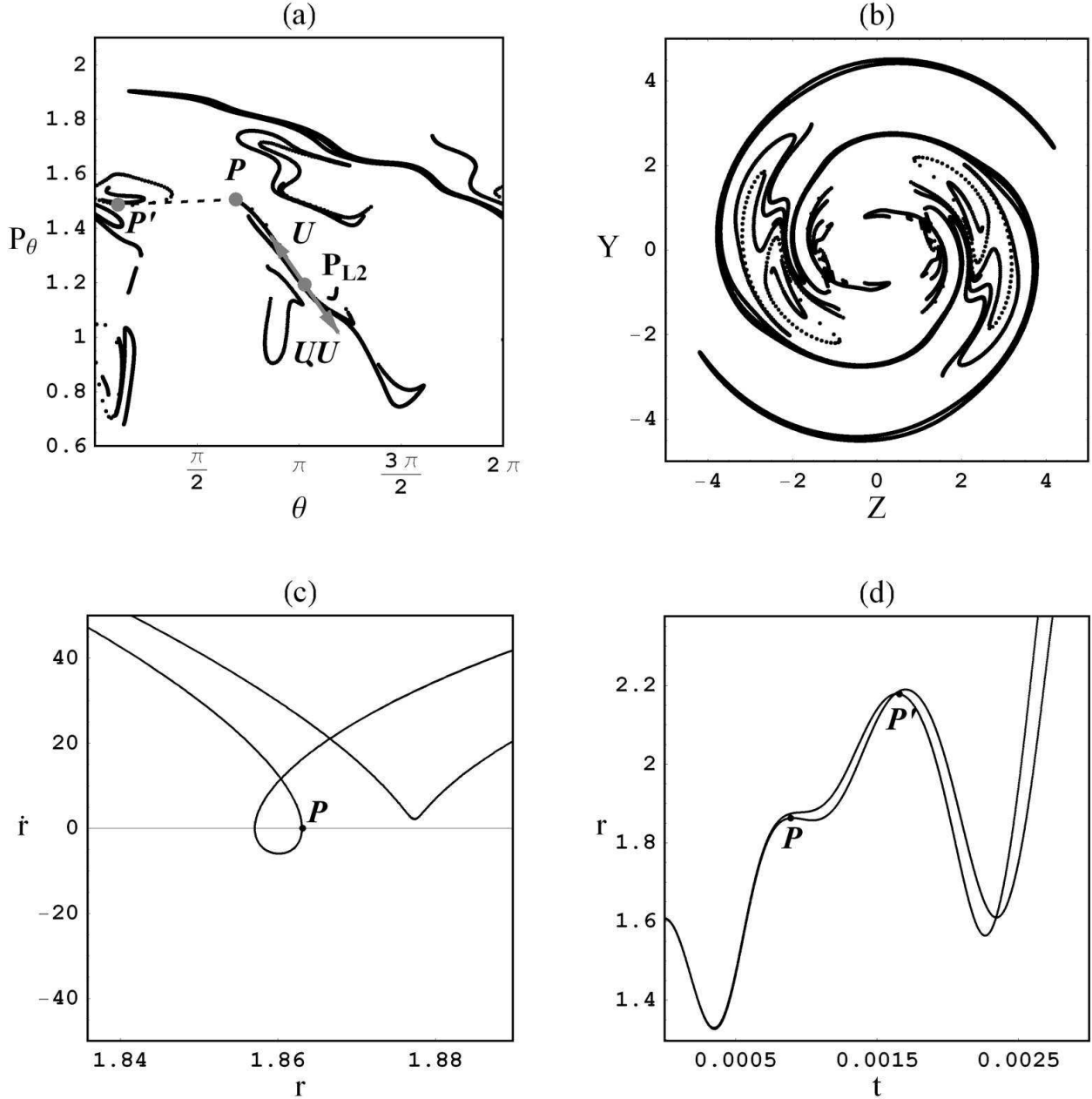}
\caption{ a) Unstable invariant manifolds of the unstable periodic
orbit $P_{L2}$ in the Poincar\'{e} section $(\theta,p_\theta)$,
$\dot{r}=0, \ddot{r}<0$ of the potential of the N-Body simulation,
at the time snapshot $t=47$. The initial directions are marked U
and UU. The initial segments along U and UU, of length
$ds=10^{-4}$, are populated with 10000 points each. The figure
shows the image of these points after six iterations of the
Poincar\'{e} map. The Jacobi constant is $E_J=-1.125\times 10^6$.
b) The manifolds of (a) projected in configuration space, together
with the symmetric manifolds with respect to the center. c) The
evolution $\dot{r}$ vs. $r$ for two orbits which give the points
of intersection $P$ and $P'$ in the Poincar\'{e} section. d) The
time evolution of $r(t)$ for the same pair of orbits.}
\label{figure7}
\end{figure}

\begin{figure}
\centering
\includegraphics[width=\textwidth]{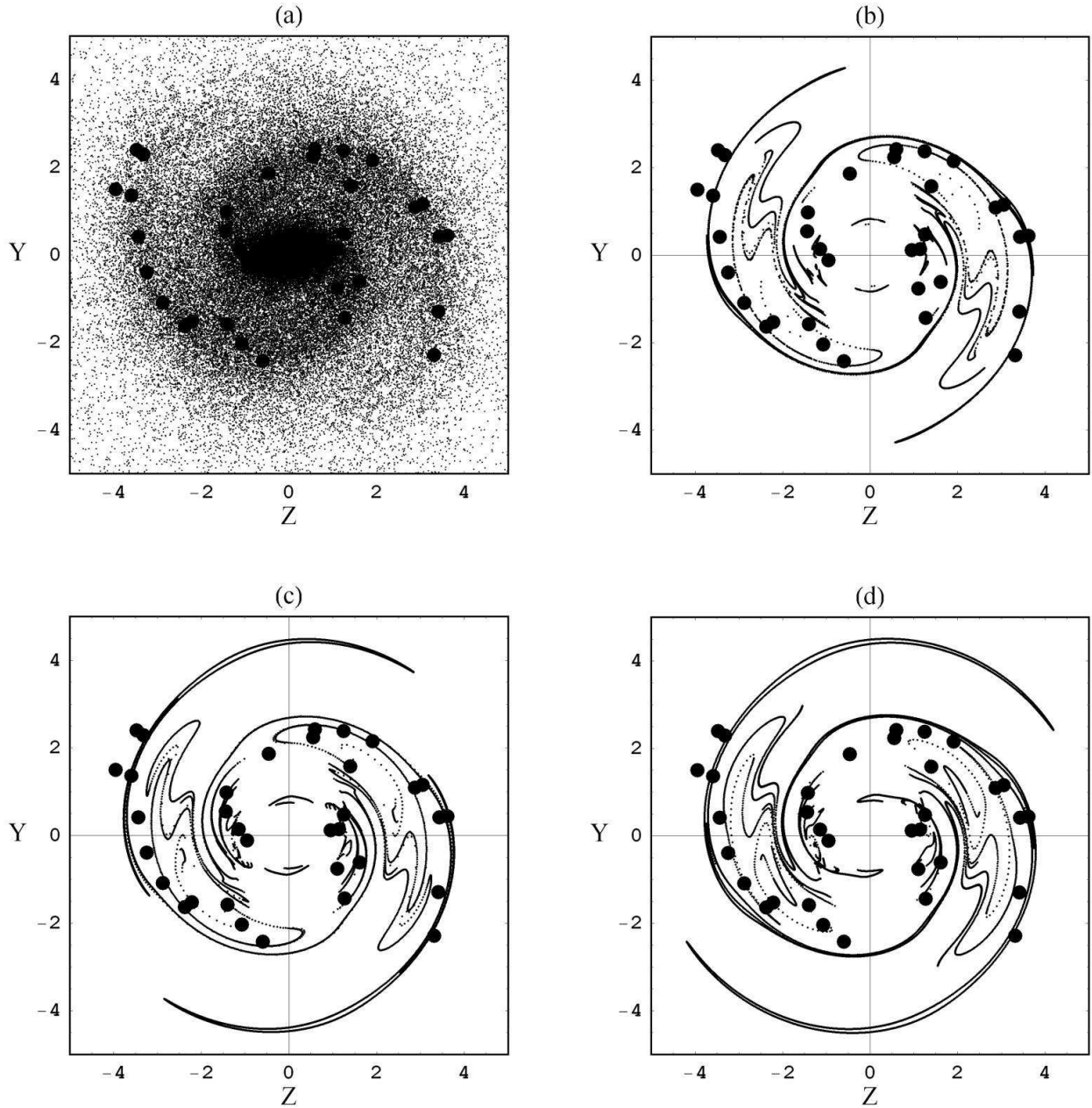}
\caption{ a) Same as in figure 5a. The form of the spiral (thick
dots) is compared with the forms of the unstable manifolds
emanating from the unstable periodic orbits $P_{L1,2}$ for three
different values of the Jacobi constant, namely b)
$E_J=-1.131\times 10^6$, c) $E_J=-1.128\times 10^6$, and d)
$E_J=-1.125\times 10^6$. } \label{figure7}
\end{figure}

\begin{figure}
\centering
\includegraphics[width=14cm]{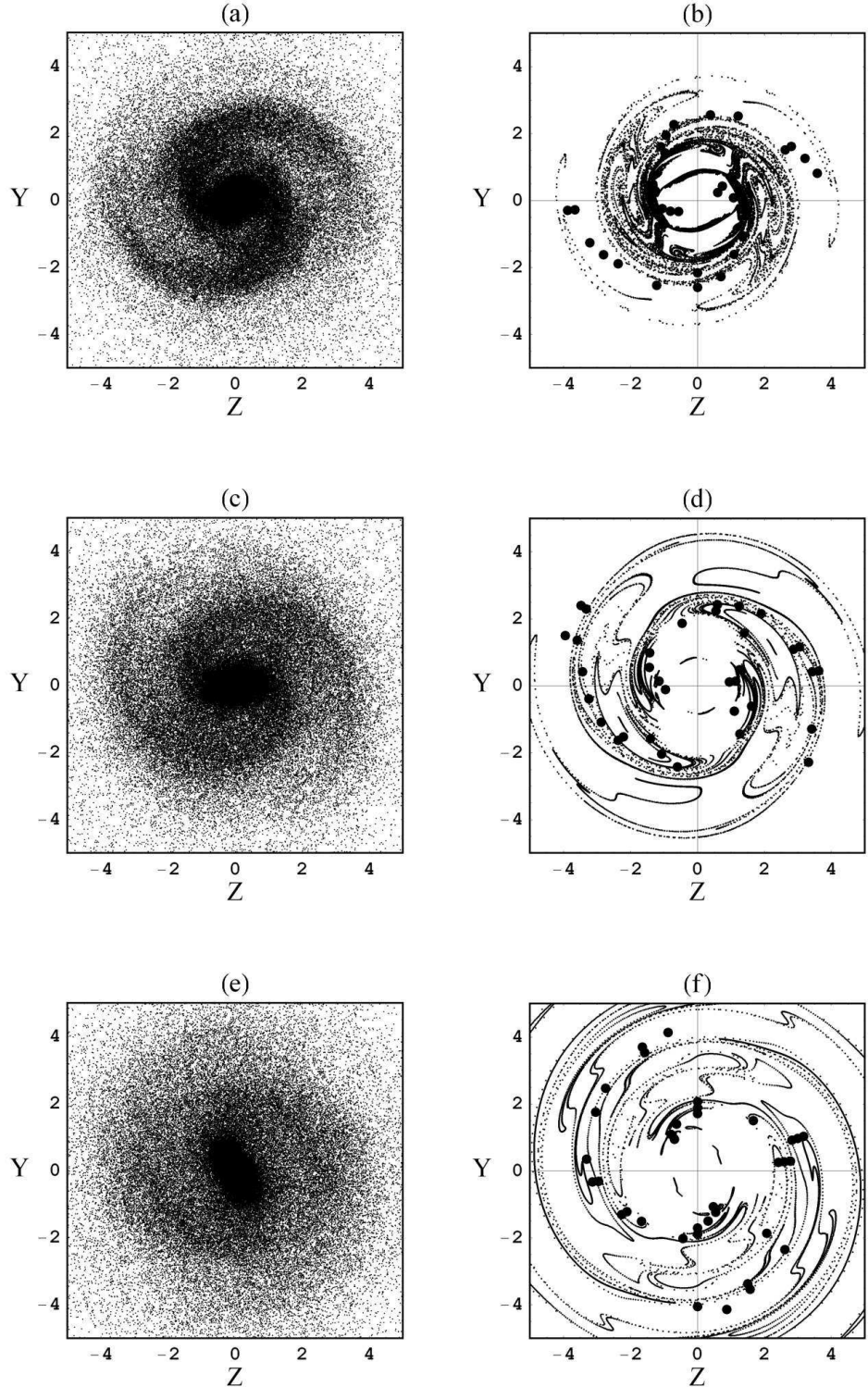}
\caption{ The projection of the N-Body system on the $(y,z)$ plane
at three different snapshots: a) $t=18$, c) $t=47$, and e) $t=74$.
The comparison of the spiral pattern with the unstable invariant
manifolds emanating from $P_{L1,2}$ is made in (b), (d) and (f)
respectively. The corresponding Jacobi constant values at which
the manifolds are calculated is $E_J=-1.25\times 10^6$ for (b),
$E_J=-1.131\times 10^6$ for (d) and $E_J=-1.085\times 10^6$ for
(f). The 8-th manifolds' iteration is shown. } \label{figure8}
\end{figure}

\begin{figure}
\centering
\includegraphics[width=14cm]{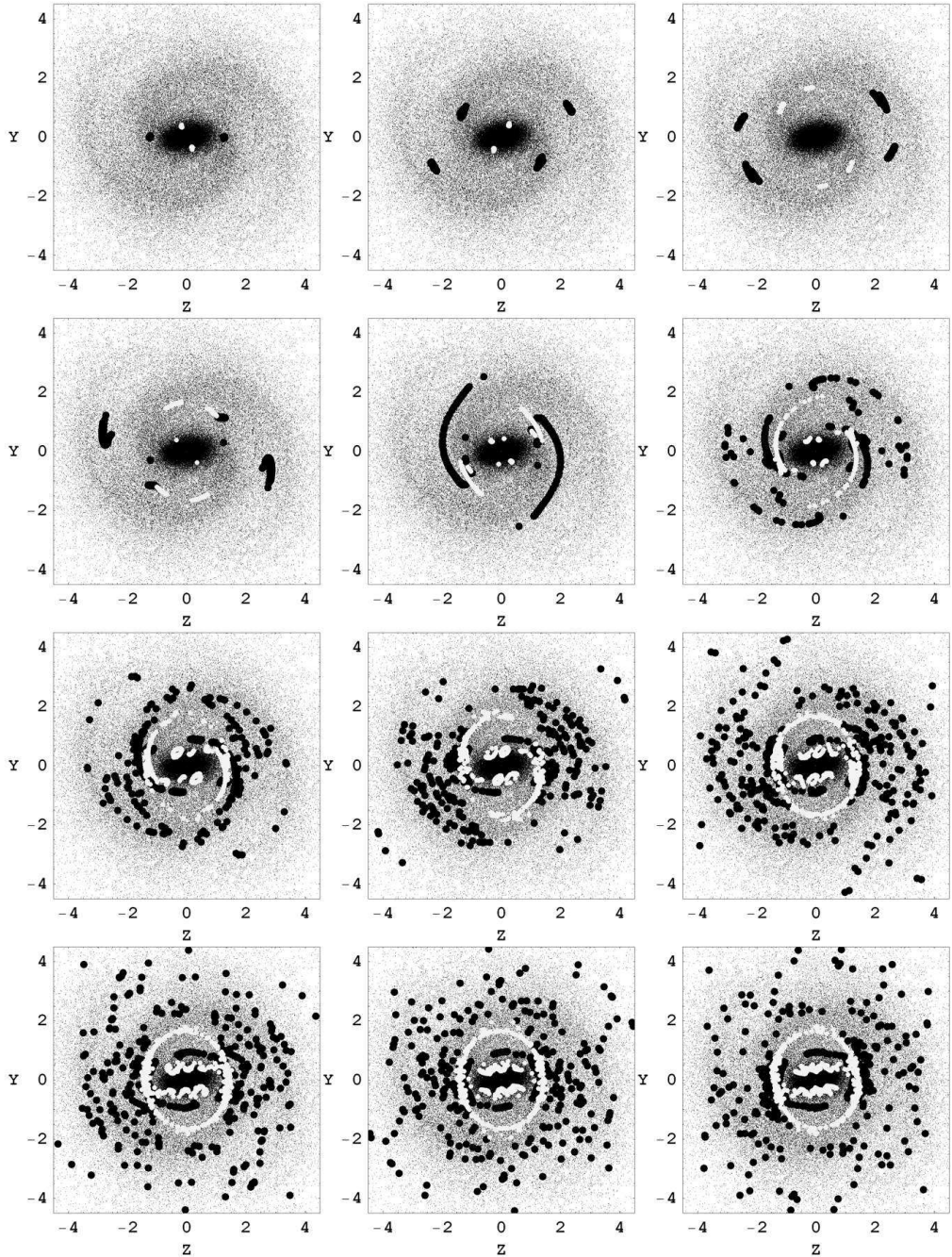}
\caption{Twelve successive Poincar\'{e} consequents of the
apocenters (thick black dots) and the pericenters (thick white
dots) of 600 orbits with initial conditions on the unstable
manifolds U emanating from the fixed points $P_{L1,2}$ (300 orbits
for each fixed point). The first panel corresponds to the
Poincar\'{e} consequent at the time when the apocenters start
leaving the neighborhood of the fixed points. The phases of the
apocenters of these orbits remain well correlated for at least 9
successive passages. The phases of the pericenters are well
correlated for much longer. The pericenters tend to form a
double-ring surrounding the bar and the stable Lagrangian points
$L_4$, $L_5$ and cooperate with apocenters to support a chaotic
shell surrounding the bar, but also an azimuthal arc of the spiral
arms near $L_1$, $L_2$, outside corotation. Spiral arms at larger
azimuthal angles and larger radii can be supported only by
apocenters.} \label{figure10}
\end{figure}

\end{document}